\documentclass[sigconf]{acmart}
\pdfoutput=1
\usepackage{booktabs} 
\usepackage[utf8]{inputenc}
\usepackage[english]{babel}

\usepackage{url}
\usepackage{hyperref} 

\usepackage{amsmath,amssymb,amsthm,wasysym,mathtools,esvect}

\usepackage[babel=true]{csquotes}
\usepackage{cleveref}

\usepackage{amsmath,amssymb,amsthm,wasysym,mathtools,esvect,dsfont}

\usepackage{graphicx}
\usepackage{color}
\makeatletter
\@ifpackageloaded{xcolor}{}{
  \usepackage[dvipsnames]{xcolor}
}

\definecolor{persianrose}{rgb}{1.0, 0.16, 0.64}
\definecolor{goldenyellow}{rgb}{1.0, 0.87, 0.0}

\makeatother

\usepackage{float}

\usepackage{tikz}
\usetikzlibrary{automata}
\usetikzlibrary{arrows}
\usetikzlibrary{shadows}
\usetikzlibrary{decorations.text}
\usetikzlibrary{circuits.logic.US} 
\usetikzlibrary{matrix}

\usepackage{pgfplots}

\usepackage{subcaption}

\usepackage{enumitem}

\usepackage{environ}

\usepackage{tabularx}
\usepackage{makecell}
\usepackage{multirow}
\usepackage{colortbl}
\usepackage{collcell}
\usepackage{threeparttable}
\usepackage{adjustbox}
\usepackage{booktabs}

\usepackage{listings}
\let\origthelstnumber\thelstnumber
\makeatletter
\newcommand*\Suppressnumber{%
  \lst@AddToHook{OnNewLine}{%
    \let\thelstnumber\relax%
     \advance\c@lstnumber-\@ne\relax%
    }
}

\newcommand*\Reactivatenumber[1]{%
  \setcounter{lstnumber}{\numexpr#1-1\relax}
  \lst@AddToHook{OnNewLine}{%
   \let\thelstnumber\origthelstnumber%
   \refstepcounter{lstnumber}
  }
}

\makeatother
\lstdefinestyle{vhdl_style}
{
    language=VHDL,
    float=!htb,
    basicstyle=\ttfamily\footnotesize,
    identifierstyle=\bfseries\color{black},
    keywordstyle=\bfseries\color{persianrose},
    stringstyle=\bfseries\color{yellow},
    commentstyle=\bfseries\color{gray},
    columns=flexible,
    frame=single,
    showspaces=false,
    showstringspaces=false,
    numberstyle=\tiny,
    stepnumber=1,
    breaklines=true,
    xrightmargin=-\fboxsep,
    backgroundcolor=\color{white},
    captionpos=t,
    mathescape,
    escapechar=\%
}

\usepackage{algorithm}
\usepackage[noend]{algpseudocode}
\algrenewcommand\algorithmicrequire{\textbf{\ \ Input:}}
\algrenewcommand\algorithmicensure{\textbf{Output:}}

\usepackage{microtype}
\usepackage{xspace}
\usepackage{pifont}
\usepackage{comment}

\usepackage{ulem}

\usepackage[nolist]{acronym}

\begin{acronym}

\acro{3DES}{Triple-DES}

\acro{AES}{Advanced Encryption Standard}
\acro{ALU}{Arithmetic Logic Unit}
\acro{API}{Application Programming Interface}
\acro{ARX}{Addition Rotation XOR}
\acro{ASIC}{Application Specific Integrated Circuit}
\acro{ASIP}{Application Specific Instruction-Set Processor}
\acro{AS}{Active Serial}

\acro{BDD}{Binary Decision Diagram}
\acro{BGL}{Boost Graph Library}
\acro{BNF}{Backus-Naur Form}
\acro{BRAM}{Block-Ram}

\acro{CBC}{Cipher Block Chaining}
\acro{CFB}{Cipher Feedback Mode}
\acro{CFG}{Control Flow Graph}
\acro{CLB}{Configurable Logic Block}
\acro{CLI}{Command Line Interface}
\acro{COFF}{Common Object File Format}
\acro{CPA}{Correlation Power Analysis}
\acro{CPU}{Central Processing Unit}
\acro{CRC}{Cyclic Redundancy Check}
\acro{CTR}{Counter}

\acro{DC}{Direct Current}
\acro{DES}{Data Encryption Standard}
\acro{DFA}{Differential Frequency Analysis}
\acro{DFT}{Discrete Fourier Transform}
\acro{DLL}{Dynamic Link Library}
\acro{DMA}{Direct Memory Access}
\acro{DNF}{Disjunctive Normal Form}
\acro{DPA}{Differential Power Analysis}
\acro{DSO}{Digital Storage Oscilloscope}
\acro{DSP}{Digital Signal Processing}
\acro{DUT}{Design Under Test}

\acro{ECB}{Electronic Code Book}
\acro{ECC}{Elliptic Curve Cryptography}
\acro{EEPROM}{Electrically Erasable Programmable Read-only Memory}
\acro{EMA}{Electromagnetic Emanation}
\acro{EM}{electro-magnetic}

\acro{FFT}{Fast Fourier Transformation}
\acro{FF}{Flip Flop}
\acro{FI}{Fault Injection}
\acro{FIR}{Finite Impulse Response}
\acro{FPGA}{Field Programmable Gate Array}
\acro{FSM}{Finite State Machine}

\acro{GUI}{Graphical User Interface}

\acro{HDL}{Hardware Description Language}
\acro{HD}{Hamming Distance}
\acro{HF}{High Frequency}
\acro{HSM}{Hardware Security Module}
\acro{HW}{Hamming Weight}

\acro{IC}{Integrated Circuit}
\acro{IO}[I/O]{Input/Output}
\acro{IOB}{Input Output Block}
\acro{IOT}[IoT]{Internet of Things}
\acro{IP}{Intellectual Property}
\acro{ISA}{Instruction Set Architecture}
\acro{IV}{Initialization Vector}

\acro{JTAG}{Joint Test Action Group}

\acro{KAT}{Known Answer Test}

\acro{LFSR}{Linear Feedback Shift Register}
\acro{LSB}{Least Significant Bit}
\acro{LUT}{Look-up table}

\acro{MAC}{Message Authentication Code}
\acro{MIPS}{Microprocessor without Interlocked Pipeline Stages}
\acro{MMIO}{Memory Mapped \acl{IO}}
\acro{MSB}{Most Significant Bit}

\acro{NASA}{National Aeronautics and Space Administration}
\acro{NSA}{National Security Agency}
\acro{NVM}{Non-Volatile Memory}

\acro{OFB}{Output Feedback Mode}
\acro{OISC}{One Instruction Set Computer}
\acro{ORAM}{Oblivious Random Access Memory}
\acro{OS}{Operating System}

\acro{PAR}{Place-and-Route}
\acro{PCB}{Printed Circuit Board}
\acro{PC}{Personal Computer}
\acro{PS}{Passive Serial}
\acro{PUF}{Physical Unclonable Function}

\acro{RISC}{Reduced Instruction Set Computer}
\acro{RNG}{Random Number Generator}
\acro{ROM}{Read-Only Memory}
\acro{ROP}{Return-oriented Programming}
\acro{RTL}{Register Transfer Level}

\acro{SCA}{Side-Channel Analysis}
\acro{SHA}{Secure Hash Algorithm}
\acro{SNR}{Signal-to-Noise Ratio}
\acro{SPA}{Simple Power Analysis}
\acro{SPI}{Serial Peripheral Interface Bus}
\acro{SRAM}{Static Random Access Memory}

\acro{UART}{Universal Asynchronous Receiver Transmitter}
\acro{UHF}{Ultra-High Frequency}

\acro{WISC}{Writeable Instruction Set Computer}

\acro{XDL}{Xilinx Description Language}
\acro{XTS}{XEX-based Tweaked-codebook with ciphertext Stealing}

\end{acronym}

\definecolor{lightcoral}{rgb}{0.94, 0.5, 0.5}
\definecolor{lightskyblue}{rgb}{0.53, 0.81, 0.98}

\newcommand{\CC}[1][]{$\text{C\hspace{-.25ex}}^{_{_{_{++}}}}\ifthenelse{\equal{#1}{}}{}{\text{\hspace{-.625ex}#1}}$\xspace}

\newcommand{\circled}[2][]{
  \tikz[baseline=(char.base)]{
    \node[shape=circle,draw,inner sep=1pt,fill=black]
    (char) {\phantom{\ifblank{#1}{#2}{#1}}};
    \node[text=white] at (char.center) {\makebox[0pt][c]{\textbf#2}};}\xspace}
\robustify{\circled}

\makeatletter
\newcommand{\removeifnextchar}[2]{%
    \begingroup
    \ltx@LocToksA{\endgroup#2}%
    \ltx@ifnextchar@nospace{#1}{%
        \def\next{\the\ltx@LocToksA}%
        \afterassignment\next
        \let\scratch= %
    }{%
        \the\ltx@LocToksA
    }%
}

\newcommand{\etal}{\protect\removeifnextchar{.}{et~al.\xspace}}

\makeatother

\newcommand{\romannumber}[1]{\uppercase\expandafter{\romannumeral#1}}

\ifdefined\algorithmautorefname

\else
  \newcommand{\algorithmautorefname}{Algorithm}
\fi 

\ifdefined\definitionautorefname

\else 
  \newcommand{\definitionautorefname}{Definition}
\fi

\addto\extrasenglish{  
  \ifdefined\algorithmautorefname
    \renewcommand{\algorithmautorefname}{Algorithm}
  \fi 
  \ifdefined\definitionautorefname
    \renewcommand{\definitionautorefname}{Definition}
  \fi

}

\newcommand{\Figure}[1]{\textcolor{blue}{\autoref{#1}}}

\newcommand{\Section}[1]{\textcolor{blue}{\autoref{#1}}}

\renewcommand{\Figure}[1]{\autoref{#1}}

\renewcommand{\Section}[1]{\autoref{#1}}

\newcommand{\HAL}{\textnormal{\textsf{HAL}}\xspace}

\newcommand*{\MinNumber}{0.0}
\newcommand*{\MidNumber}{0.76}
\newcommand*{\MaxNumber}{1.0}

\newcommand{\ApplyGradient}[1]{
  \ifdim #1 pt > \MidNumber pt
    \pgfmathsetmacro{\PercentColor}{max(min(100.0*(#1 - \MidNumber)/(\MaxNumber-\MidNumber),100.0),0.00)} %
    \colorbox{green!\PercentColor!yellow}{#1}
  \else
    \pgfmathsetmacro{\PercentColor}{max(min(100.0*(\MidNumber - #1)/(\MidNumber-\MinNumber),100.0),0.00)} %
    \colorbox{red!\PercentColor!yellow}{#1}
  \fi
}

\newcolumntype{R}{>{\collectcell\ApplyGradient}c<{\endcollectcell}}

\setcopyright{acmcopyright}

\usepackage{lmodern}

\acmDOI{10.475/123_4}

\acmISBN{123-4567-24-567/08/06}

\acmConference[ASP-DAC 2019]{ACM Asia South Pacific Design Automation Conference}{Jan. 2019}{Tokyo, Japan}
\acmYear{2019}
\copyrightyear{2019}

\acmArticle{4}
\acmPrice{15.00}

\begin{document}
	\title{Towards Cognitive Obfuscation}
	\subtitle{Impeding Hardware Reverse Engineering Based on Psychological Insights}

	\author{Carina Wiesen}
    \authornote{Educational Research Institute, Ruhr-Universität Bochum}
	\orcid{1234-5678-9012}
	\affiliation{%
		\institution{Ruhr-Universität Bochum}
		\streetaddress{Universitätsstr. 150, 44801 Bochum}
		\city{Bochum}
		\state{Germany}
		\postcode{44801}
	}
	\email{carina.wiesen@rub.de}

	\author{Nils Albartus}
	\authornote{Horst Görtz Institute for IT Security, Ruhr-Universität Bochum}
	\orcid{1234-5678-9012}
	\affiliation{%
		\institution{Ruhr-Universität Bochum}
		\streetaddress{Universitätsstr. 150, 44801 Bochum}
		\city{Bochum}
		\state{Germany}
		\postcode{44801}
	}
	\email{nils.albartus@rub.de}

	\author{Max Hoffmann}
    \authornotemark[2]
		\affiliation{%
		\institution{Ruhr-Universität Bochum}
		\streetaddress{Universitätsstr. 150, 44801 Bochum}
		\city{Bochum}
		\state{Germany}
		\postcode{44801}
	}
	\email{max.hoffmann@rub.de}

	\author{Steffen Becker}
    \authornotemark[2]
	\affiliation{%
		\institution{Ruhr-Universität Bochum}
		\streetaddress{Universitätsstr. 150, 44801 Bochum}
		\city{Bochum}
		\state{Germany}
		\postcode{44801}
	}
	\email{steffen.becker@rub.de}

	\author{Sebastian Wallat}
	\affiliation{%
		\institution{University of Massachusetts}
		\streetaddress{Universitätsstr. 150, 44801 Bochum}
		\city{Amherst}
		\state{Massachusetts}
		\postcode{44801}
	}
	\email{swallat@umass.edu}

	\author{Marc Fyrbiak}
    \authornotemark[2]
	\affiliation{%
		\institution{Ruhr-Universität Bochum}
		\streetaddress{Universitätsstr. 150, 44801 Bochum}
		\city{Bochum}
		\state{Germany}
		\postcode{44801}
	}
	\email{marc.fyrbiak@rub.de}

	\author{Nikol Rummel}
    \authornotemark[1]
    \authornotemark[2]
	\affiliation{%
		\institution{Ruhr-Universität Bochum}
		\streetaddress{Universitätsstr. 150, 44801 Bochum}
		\city{Bochum}
		\state{Germany}
		\postcode{44801}
	}
	\email{nikol.rummel@rub.de}

	\author{Christof Paar}
    \authornotemark[2]
	\affiliation{%
		\institution{Ruhr-Universität Bochum}
		\streetaddress{Universitätsstr. 150, 44801 Bochum}
		\city{Bochum}
		\state{Germany}
		\postcode{44801}
	}
	\email{christof.paar@rub.de}

	\renewcommand{\shortauthors}{C. Wiesen et al.}

	\begin{abstract}
		In contrast to software reverse engineering, there are hardly any tools available that support hardware reversing.
		Therefore, the reversing process is conducted by human analysts combining several complex semi-automated steps.
		However, countermeasures against reversing are evaluated solely against mathematical models.
		Our research goal is the establishment of cognitive obfuscation based on the exploration of underlying psychological processes.
		We aim to identify problems which are hard to solve for human analysts and derive novel quantification metrics, thus enabling stronger obfuscation techniques.
	\end{abstract}

	\maketitle



\section{Introduction}
Hardware components form the root of trust in virtually any computing system.
In order to detect fabrication faults, copyright infringements, counterfeit products, or malicious manipulations hardware reverse engineering is usually the tool-of-choice \cite{guin_counterfeit_2014, ieee:2014:bhunia, ivsw:2017:fyrbiak}.
While hardware reverse engineering is a highly complex and universal tool for legitimate purposes, it can also be employed with illegitimate intentions, undermining the integrity of ICs via piracy, subsequent weakening of security functions, or insertion of hardware Trojans \cite{jetc:2016:quadir, ivsw:2017:fyrbiak}.
In particular, \ac{IP} piracy has become a major concern for the semiconductor industry which causes losses in the range of several billion dollars \cite{guin_counterfeit_2014}.
Governments and armed forces classify hardware reverse engineering as a major security threat, since malicious manipulations of hardware components in mission-critical systems can even have lethal consequences \cite{ivsw:2017:fyrbiak}.

Due to the serious threats posed by attacks based on hardware reverse engineering, strong countermeasures are indispensable.
The security of most existing obfuscation techniques is assessed exclusively based on technical measures.
However, the process of hardware reverse engineering cannot be fully automated yet and the lack of holistic tools forces human analysts to combine several semi-automated steps \cite{ches:2009:torrance}.
Accordingly, cognitive processes and strategies applied by humans in the context of hardware reverse engineering must be taken into account for the development of sound countermeasures.
In this work, we describe a novel approach of developing \textit{cognitive obfuscation} methods considering both technical and human factors during the hardware reverse engineering processes.

\textbf{Goal and Contribution.} Our work is motivated by the lack of metrics to precisely capture the strength of defenses against hardware reverse engineering.
Since technical and human factors are inseparably connected in hardware reverse engineering processes, the thorough understanding of the human factor is essential to develop cognitive obfuscation methods.
To the best of our knowledge we are the first to define cognitive obfuscation as the approach of impeding hardware reverse engineering based on the analysis of psychological processes.
In summary our contributions are:
\begin{itemize}
	\item \textbf{Exploration of Human Factors.} We define our psychological research design to analyze the underlying cognitive processes of hardware reverse engineering in detail.
	Additionally, we emphasize how reverse engineering can be formulated as a psychological concept of problem solving and identify which insights will perspectively help to develop sound cognitive obfuscation methods.

	\item \textbf{Improved Hardware Reverse Engineering Tool \HAL.} We significantly advance the development of our hardware reverse engineering software \HAL \cite{tdsc:2018:fyrbiak} to adequately capture human behavior as a foundation for this research.

	\item \textbf{Cognitive Obfuscation Methods.} We derive the necessity for novel obfuscation methods based on psychological insights from the inseparable connection of technical and human factors in hardware reverse engineering.
\end{itemize}



\section{Technical Background}
In the following section we define the term hardware reverse engineering and specify gate-level netlist reverse engineering as our main research topic.
Additionally, we provide a brief overview on existing obfuscation methods impeding gate-level netlist reverse engineering and emphasize the significance of our novel approach leading to cognitive obfuscation methods.

\subsection{Hardware Reverse Engineering}
Reverse engineering is commonly described as the process of retrieving knowledge or information from anything man-made in order to understand its inner structure and workings \cite{rekoff_reverse_1985}.
In the context of computer security, reverse engineering is often associated with the analysis of binary programs \cite{willems_reverse_2012} or hardware designs \cite{forte_hardware_2017}.

Reverse engineering in the field of hardware security is conducted for legitimate as well as illegitimate purposes \cite{ivsw:2017:fyrbiak}.
On the one hand, security engineers are often forced to conduct reverse engineering to identify counterfeited hardware, security vulnerabilities, or malicious manipulations like hardware Trojans \cite{ivsw:2017:wallat}.
On the other hand, hardware reverse engineering is also associated with illegitimate actions such as the analysis of hardware designs on various levels to understand their functionality in order to commit \ac{IP} infringement, weaken security functions, or to inject Trojans \cite{ivsw:2017:fyrbiak}.

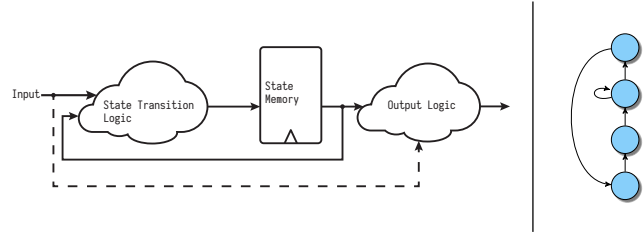
\begin{figure}[htb]
	\centering
	\resizebox{\linewidth}{!}{%
		\begin{tikzpicture}
		\tikzstyle{every state}=[fill=white,draw=black,text=black,inner sep=0pt,minimum size=17pt,circular drop shadow]

		\node  at (2.6,2.5) {\includegraphics[scale=0.58]{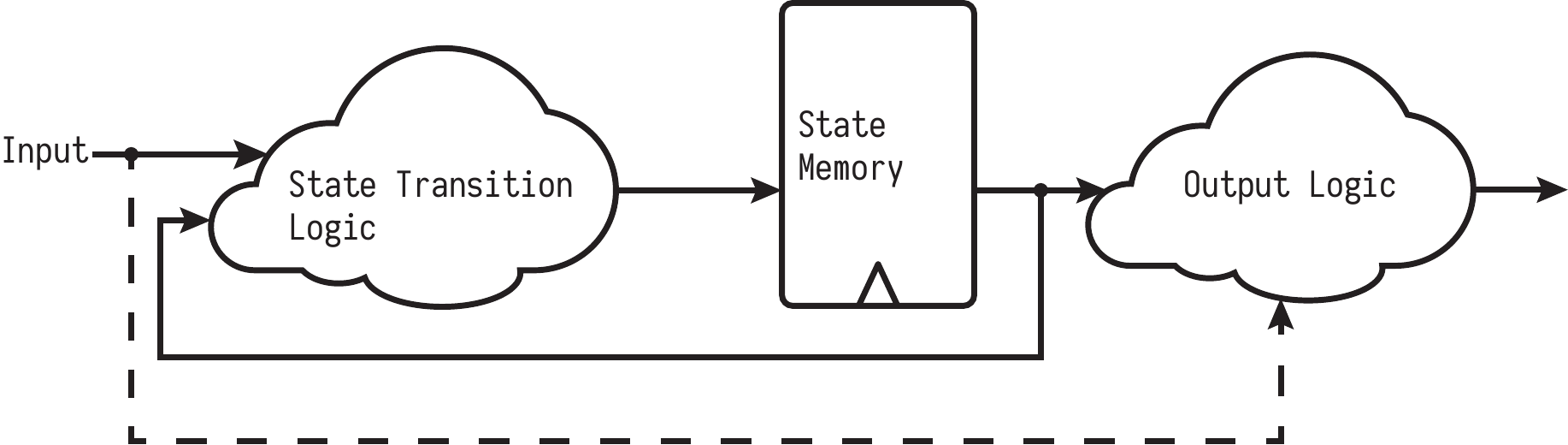}};

		\draw [thick] (8.5,0) -- (8.5,5);

		\node[state,fill=lightskyblue,scale=1] (s0) at (10.5, 1) {};
		\node[state,fill=lightskyblue,scale=1] (s1) at (10.5, 2) {};
		\node[state,fill=lightskyblue,scale=1] (s2) at (10.5, 3) {};
		\node[state,fill=lightskyblue,scale=1] (s3) at (10.5, 4) {};
		\path [->,>=stealth'] (s0) edge (s1) (s1) edge (s2) (s2) edge (s3) (s2) edge [loop left] (s2) (s3) edge [bend angle=85, bend right] (s0);

		\end{tikzpicture}

	}
	\caption{Schematic of an \ac{FSM} circuit in hardware}
	\label{co::fig::fsm_in_hardware}
\end{figure}

A valuable reference point and target of hardware reverse engineers are  \acp{FSM}. Building the control logic, \acp{FSM} form an integral part of virtually any hardware design. Due to its small circuitry and unique structure - consisting of a feedback path, which connects the state transition logic and state memory (see \Figure{co::fig::fsm_in_hardware}) - the FSM can be \textit{easily} identified in the \textit{sea of gates}~\cite{ches:2018:fyrbiak}.

Another valuable target for reverse engineers are implementations of cryptographic ciphers -- like the Advanced Encryption Standard (AES) -- which are often implemented in hardware to accelerate encryption and decryption processes.
Reverse engineering of such implementations can reveal weaknesses or even hard-coded keys in the worst-case.
Furthermore, the acquired knowledge can serve as a starting point to insert a hardware Trojan into the given structure.

\paragraph{Gate-level Netlist Reverse Engineering}
A gate-level netlist is a representation of a set of logic gates of a particular gate library together with their interconnections \cite{weste_cmos_2011}.
In this work we will focus on \ac{FPGA} gate-level netlist reverse engineering.
In \ac{FPGA} netlists combinational logic is usually implemented with Look-Up-Tables (LUTs) and Multiplexers (MUX), while sequential logic is realized with Flip-Flops (FFs).

During synthesis, which transforms a high-level description of the circuit into the gate-level netlist, important human-readable information is lost, namely (1) meaningful descriptive labels, (2) boundaries of implemented modules, and (3) module hierarchies.
The absence of this information drastically complicates the reverse engineering process and forces human analysts to conduct non-automatable sense-making processes \cite{ivsw:2017:fyrbiak}.

\subsection{Obfuscation Methods}
\label{asp-dac:chap:obfuscation}
Obfuscation itself describes a transformation which obstructs high-level information without changing functionality.
The goal of obfuscation is to impede the reverse engineering processes.

Hardware designs can be separated in data path and control path.
While there is a lack of data path obfuscation, control path obfuscation has received manifold contributions by the scientific community.
Many works, e.g., \cite{tcad:2009:chakraborty}, \cite{tcad:2017:dofe}, \cite{usenix:2007:alkabani}, and \cite{acm:2013:desai}, introduced techniques which modify the behavior of an \ac{FSM}.
Simplified, dummy states are added and the amount of possible transitions is heavily increased, thus \textit{obfuscating} the original behavior of the \ac{FSM}.
The proposed security analyses of these obfuscation methods almost solely consider technical or mathematical attacks and often prove security in the respective models.
However, Fyrbiak et al. presented a variety of attacks against such techniques \cite{ches:2018:fyrbiak}.
These attacks were successful, because the obfuscated parts of the \acp{FSM} could be semi-automatically distinguished from original parts.
For these distinguishers, human sense-making processes and goal-oriented strategies were indispensable to differentiate both parts and eventually led to practical attacks against presumably secure techniques.

We emphasize, that security assessments and proofs of the aforementioned techniques were solely based on technical models.
Incorporating knowledge about the underlying psychological processes faced by human analysts during the reverse engineering of hardware designs might have directly uncovered the anchor points used by Fyrbiak et al. in \cite{ches:2018:fyrbiak}.

An holistic overview of recent metrics and obfuscation methods for hardware can be found in~\cite{book:hardware_obfuscation:chapter1}.

\subsection{Need for Cognitive Obfuscation Methods to Impede Hardware Reverse Engineering}
Naturally, the absence of high-level information in gate-level netlists requires human analysts to employ sense-making processes during reverse engineering.
Hence, human strategies and cognitive abilities influence the success of the reverse engineering process.
If we want to develop stronger obfuscation methods it is essential to understand how human analysts are reversing a hardware design.
By acquiring a deeper understanding of used strategies or hard problems for a reverse engineer we can derive new metrics to assess the strength of an obfuscation scheme. 
Eventually, these metrics can lead to stronger approaches for hardware obfuscation based on cognitive insights.



\section{Psychological Background}
Human factors have not been taken into account for the development of obfuscation methods, yet.
This is surprising, since successful reverse engineering of netlists requires human analysts to employ their cognitive abilities to design strategies manually.
Hence, human factors and underlying psychological processes play a crucial role in hardware reverse engineering and should be included in effective obfuscation methods.
To the best of our knowledge we are one of the first to explore relevant human factors and underlying psychological processes for the development of cognitive obfuscation methods.
In this section we provide the psychological background required to describe and analyze human behavior in hardware reverse engineering.
Our research is based on the theory of reverse engineering and its application to Boolean systems by Lee \etal \cite{lee_theory_2013} which will be presented in the following.

\subsection{Reverse Engineering - A Special Kind of Problem Solving?}

Lee and Johnson-Laird \cite{lee_theory_2013} explored underlying psychological processes of reverse engineering and defined it as a special kind of problem solving. Problem solving comprises psychological competencies which involve goal-directed thinking and actions in situations, in which the problem solver does not operate routinely \cite{greiff_problemlosen_2014}. Accordingly, a problem exists when a person does not have the relevant knowledge to produce an immediate solution.

In psychological research, two kinds of problem solving are commonly distinguished \cite{funke_problemlosendes_2003}, \cite{leutner_analytische_2012}. First in analytical problems all relevant information to successfully solve the problem are included. A common example of an analytical problem is the Tower of Hanoi \cite{kotovsky_why_1985}. The challenge of analytical problems are the analysis and sequential utilization of given information \cite{greiff_problemlosen_2014}. The second type of problems are defined as complex problems which do not specify information to solve the problem. In complex problem situation the problem solver is asked to manipulate information to solve the problem. Complex problems combine different characteristics like non-transparency of information, dependencies between components of the problem, dynamics, or multiple goals \cite{greiff_problemlosen_2014}.

\begin{figure}[!htb]
	\centering
	\includegraphics[width=\linewidth]{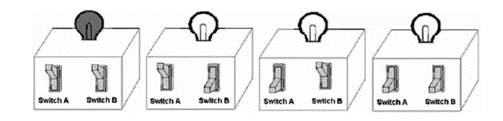}
	\caption{Example of Reverse Engineering Task (Lee \& Johnson-Laird, 2013) \cite{lee_theory_2013}}
	\label{co::fig::lamps}
\end{figure}

\begin{figure}[!htb]
	\centering
	\includegraphics[width=0.4\linewidth]{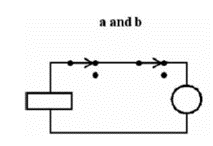}
	\caption{Solution of Reverse Engineering Task in Figure 1 (Lee \& Johnson-Laird, 2013) \cite{lee_theory_2013}
}
	\label{co::fig::solution}
\end{figure}

Lee \& Johnson-Laird \cite{lee_theory_2013} argue that reverse engineering involves both attributes of analytical and complex problem solving. Hence, problem solver in reverse engineering divide a given system (like a netlist) in subsystems which is a common strategy in analytical problem solving. Additionally, dependencies between single components influence the problem solving process in reverse engineering \cite{lee_theory_2013}. Dependencies and their influence on the problem solving process are typically allocated to complex problem solving processes \cite{lee_theory_2013}. By combining characteristics of both analytical and complex problem solving, reverse engineering can be defined as a special kind of problem solving.
In the context of the theory of reverse engineering, Lee and Johnson-Laird \cite{lee_theory_2013} conducted five experiments consisting of high-level reverse engineering tasks (see \Figure{co::fig::lamps} and \Figure{co::fig::solution}).

Participants were asked to draw the underlying circuit which represents the light as on. Both congruent (\texttt{AND}, \texttt{OR}, \texttt{XOR}) and incongruent (\texttt{NAND}, \texttt{NOR}, \texttt{XNOR}) were depicted in the single tasks. Incongruent tasks were harder to solve than congruent tasks. Additionally, the difficulty of reverse engineering was significantly influenced by the number of dependencies between single components which influence the output of a system. Summarized, these results can be seen as a starting point for further research. Nevertheless, it is essential to transfer the experimental setup in the field of hardware reverse engineering to produce higher validity of the results. By the exploration of human processes in realistic hardware reverse engineering tasks, we aim to find more specific answers about how human analysts are reversing a gate-level netlist and which parameters influence the difficulty of solving hardware reverse engineering tasks in realistic scenarios. Knowledge about parameters which force engineers to make mistakes while reversing a chip can be the foundation for the development of cognitive obfuscation methods. In the following we present our novel approach of the exploration of underlying human processes in hardware reverse engineering and how we plan to use insights about human factors to derive novel cognitive obfuscation methods to impede hardware reverse engineering.

\subsection{A Novel Approach: Cognitive Obfuscation}
To the best of our knowledge we are one of the first to derive novel obfuscation methods impeding hardware reverse engineering based on psychological insights. Therefore, we aim to achieve a deeper understanding about underlying psychological processes and how these processes are influenced by human factors. For developing novel obfuscation methods, we first have to explore human processes and to identify the difficulties human analysts are facing during the process of reversing a netlist. In the next step, we can identify and explain common difficulties and recurring mistakes of human analysts with different levels of expertise. We can use these insights to develop cognitive obfuscation methods which force engineers to make these kind of mistakes. Furthermore, we are able to develop cognitive obfuscation methods based on prior results about difficulties and learning processes to test them with human analysts. This enables us to quantify different approaches of cognitive obfuscation and to derive metrics about the strength of each individual approach.

Due to the lack of prior research findings about human processes in realistic hardware reverse engineering scenarios, we formulated the following research questions.
They seek to answer if engineers are reversing hardware more efficiently with growing experience, what sort of difficulties occur over time, if these difficulties can be avoided over time, and what other cognitive factors influence the problem solving processes of hardware reverse engineering.

\paragraph{RQ1:} Do participants get more efficient in solving hardware reverse engineering tasks with growing experience?
\paragraph{RQ2:} Which difficulties occur during the learning processes?
\paragraph{RQ3:} Do the same difficulties occur repeatedly during the learning processes?
\paragraph{RQ4:} How much time do participants need to solve difficulties in the learning processes?
\paragraph{RQ5:} Which cognitive factors play a role for learning hardware reverse engineering?
\\


\section{Method}
In this section we present our research methods, materials, and designs. To answer the research questions above, we employ both longitudinal and cross-sectional research methods for analyzing the behavior of human analysts on different levels of expertise.

\subsection{Psychological Research Methods}
Longitudinal studies are common psychological research methods and often conducted to repeatedly observe behavioral changes of individual participants over a period of time \cite{shadish_experimental_2001}. In comparison to cross-sectional studies, longitudinal studies analyze behavioral changes within one person. The analysis of longitudinal case studies will focus on individual strategies in learning processes and in solving difficulties while working on hardware reverse engineering tasks. 

Cross-sectional studies are also commonly used in psychological research and aim to compare behavior between different study participants \cite{carlson_psychology:_2010}. Cross-sectional studies will focus on the comparison of used strategies, difficulties between different human analysts, and influences of cognitive factors as well as different levels of expertise.

By combining these two research methods we intend to receive a holistic understanding of underlying psychological processes of humans conducting hardware reverse engineering. Therefore, we analyze behavioral changes in individual human analysts over a period of time on the one hand and compare behavior between different analysts on the other hand. In the first step, we conduct two longitudinal studies by analyzing case studies of individual students. In the next section we outline the study procedures and describe the participants of the two longitudinal studies.

\begin{figure*}[!htb]
	\centering
	\includegraphics[width=\linewidth]{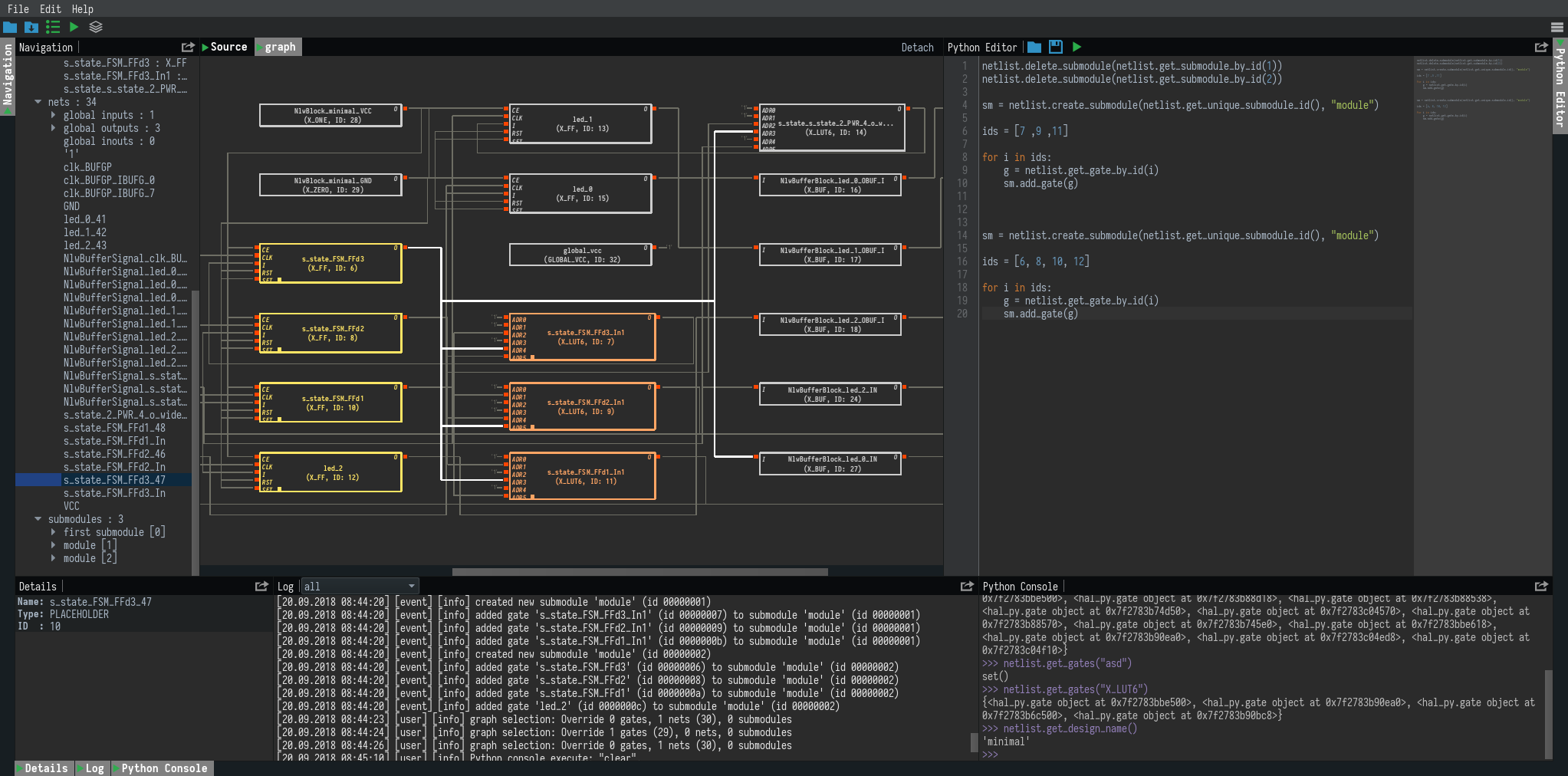}
	\caption{The completely revised GUI of \HAL}
	\label{co::fig::hal_gui}
\end{figure*}

\subsection{Longitudinal Studies: Participants \& Procedures}
Both studies are integrated in a novel educational hardware reverse engineering lab course held in winter term 2018/2019 at Ruhr-Universität Bochum (Germany) and at the University of Massachusetts Amherst (USA) \cite{tale:2018:wiesen}. The participants were novice IT security students with backgrounds in computer security, electrical engineering, or computer science. This lab course consists of two parts: the lecture phase and the practical phase. During the lecture phase students acquire relevant knowledge in Boolean algebra, general knowledge about hardware - especially \acp{FPGA} and ASICs - Hardware Description Languages (HDLs) like Verilog and VHDL, attack strategies, and obfuscation techniques. In the second phase, students can apply their newly acquired theoretical knowledge by working individually on four realistic hardware reverse engineering projects. The study is included in the practical phase and the following research data will be generated.

\paragraph{Behavioral Data.} Behavioral data of reverse engineering processes will be analyzed based on log files. To enable participants to work on hardware reverse engineering tasks, we developed four hardware reverse engineering projects as educational and study material. In all projects (\Section{sec:projects}) practical hardware reverse engineering tasks are included which students had to solve by working with our hardware reverse engineering software \HAL (\Section{co::sec::hal_improvements}). While working with \HAL, every single action is recorded automatically and will be saved as a text file including timestamps and actions, e.g. python commands, selecting relevant components in the netlist, or retrieving logic functions. Based on this data, we aim to analyze if students get more efficient in solving hardware reverse engineering tasks, what sort of difficulties or mistakes occurred and how much time they needed to solve them. Please note that the data collection phase has not been finished yet and first results will be presented in more detail at the conference.

\paragraph{Questionnaires.} Additionally, participants are asked to answer a number of questionnaires regarding information about socio-demographics, intelligence \cite{wechsler_wais-iv:_2008}, motivation \cite{vollmeyer_motivational_2006}, prior experiences in hardware reverse engineering and related topics, task difficulty \cite{bratfisch_perceived_1972}, and mental effort \cite{paas_training_1992}. Based on this data we aim to analyze which cognitive and motivational factors or levels of prior knowledge influence the success in hardware reverse engineering.

In the following we provide deeper insights about our hardware reverse engineering software \HAL and the single projects included in the study.


\subsection{\HAL}
\label{co::sec::hal_improvements}
\begin{figure}[!htb]
	\centering
	\includegraphics[width=\linewidth]{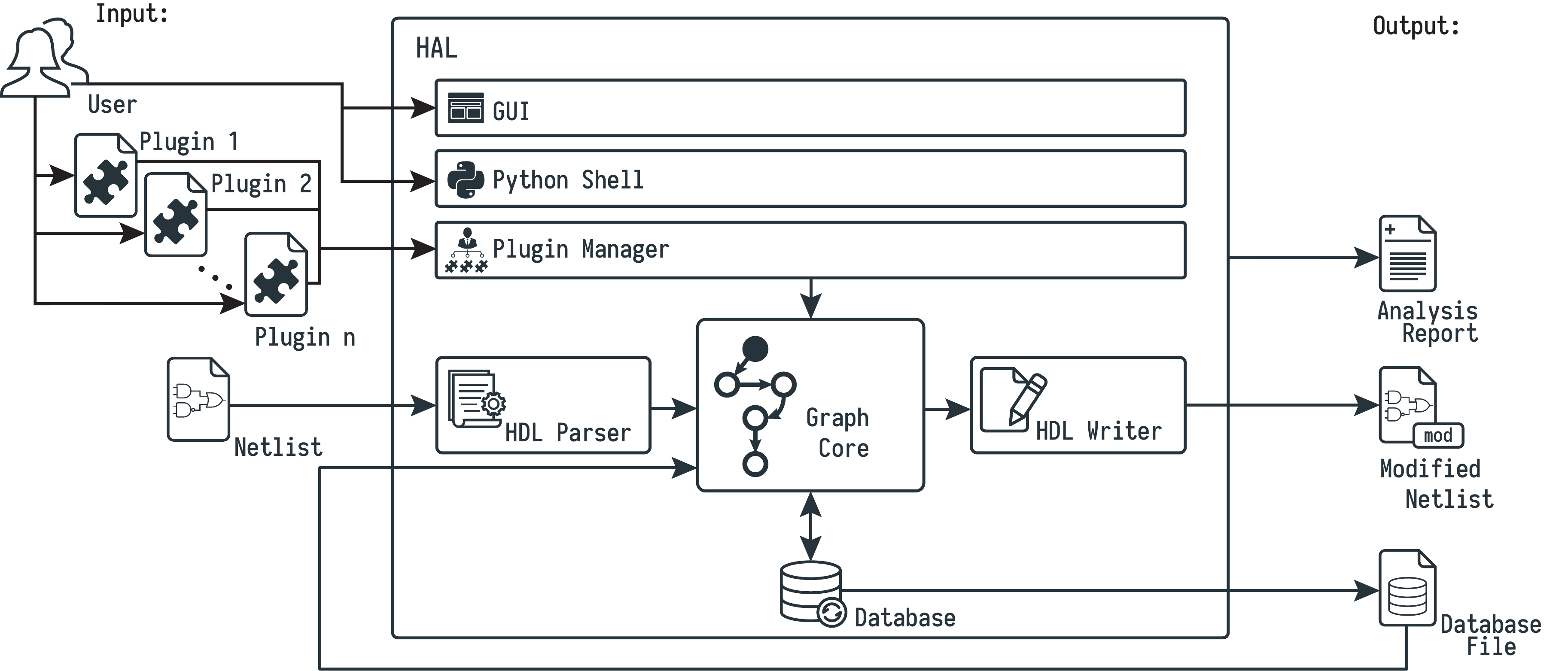}
	\caption{Overview of the original \HAL architecture from \cite{tdsc:2018:fyrbiak}}
	\label{co::fig::hal_structure}

\end{figure}

Fyrbiak et al. introduced \HAL \cite{tdsc:2018:fyrbiak}, a framework to enable (semi-)automated hardware reverse engineering.
The structure of \HAL is shown in Figure \ref{co::fig::hal_structure}.
\HAL provides a solid foundation enabling reverse engineers to implement custom plugins for netlist analysis that operate on a graph-based abstraction.
This abstraction allows for inspection of arbitrary netlists independent of specific gate-libraries or target architectures.

At the time of publication, \HAL was mainly a command-line tool which enabled implementation of arbitrary algorithms through its plugin system.
In order to yield verifiable results for psychological analysis we notably improved the GUI of \HAL.
The design goals were (1) reproducibility, (2) granularity, and (3) more user-friendly.
Therefore, we extended the GUI with an event system, a submodule partitioning system, and connected the underlying graph model to a GUI-integrated python context.
Figure \ref{co::fig::hal_gui} shows the new GUI of \HAL.
In multiple dockable widgets a variety of tools are available to the analyst, e.g., a graphical representation of the netlist, detailed information for selected elements, a rich logging window, a python shell and a python code editor.

\paragraph{Event System}
The new event system allows for exact tracing of user actions.
Regardless of the inducing source (user, plugin, or python code executed live), the core fires detailed events to notify both, GUI and logger, about changes on the data model.
While such an event system is generally used in GUI and game design, we explicitly designed this system to create information which can be used by a psychological analyst to trace the user's actions.
The main functionality of the former GUI was to allow for interactive netlist exploration.
However, it only displayed an immutable static netlist and was not connected to the plugin system.
This is only of limited benefit to a reverse engineer who wants to use both, visual inspection and code/plugins, especially when facing large designs.
With the new event system the GUI can react to changes in the data model, as well as associated data, regardless of the source, and display the new information accordingly.

\paragraph{Submodule Partitioning System}
While textual output is a primary method of reporting analysis results, \HAL allowed a reverse engineer to apply a color to a net or gate in order to create visual groups.
However, these colors were isolated GUI features and could not be created or accessed in plugins or from python code.

We replaced the isolated coloring system with a new system that allows the user to group identified gates or nets into logical modules, e.g., an identified transceiver circuit.
These so called \textit{submodules} are persistent and can be created and accessed via code and interaction with the GUI.
To preserve the visual separation, each submodule is rendered in a separate color to allow for intuitive visually distinction.
Since a gate or net can belong to multiple submodules, colored squares on each gate represent all submodules a gate is a member of.
In addition the hierarchy of submodules can be changed to determine which submodule's color has higher priority during rendering.
We plan to also include collapsing of submodules into single abstract elements in the future.

\paragraph{Python Integration}
The powerful plugin system of \HAL enables implementation of arbitrary algorithms in C++.
However, a reverse engineer often has to manually explore the adjacent area of identified modules.
Going through the rewrite/compile/debug cycle of C++ development can make this process tedious and slow.
Therefore, the original version of \HAL featured a python console, i.e., the user could instead of executing a plugin switch to a python console.
The intuitive and, more importantly, interpreted nature of python allows for dynamic execution of \HAL's features, circumventing the necessity to develop plugins for every small step.

The downside was a clear separation of \HAL runs between python console usage, GUI invocation, and plugin calls.
We have integrated the python console into the GUI together with a python code editor, allowing for interactive development and execution of scripts (cf. Figure \ref{co::fig::hal_gui}, right side) without the necessity to restart \HAL for each use-case.
To further improve interoperability, the python context is connected to the underlying graph model of the netlist and all operations are directly executed on the data model.
This allows for interleaved usage of C++ plugins, python scripts, and user actions in the GUI.

\label{sec:projects}
\subsection{Educational \& Research Material: Hardware Reverse Engineering Projects}
With our enhancements to \HAL, we were able to accompany our lectures with practical reverse engineering.
At the same time, we created a platform which enables reproducible evaluation of human processes by means of several reverse engineering projects.
The purpose of these projects is twofold.
On the one hand the projects were developed to accomplish educational goals teaching students how to practically apply various reverse engineering techniques.
On the other hand they had to be implemented as research materials to analyze human behavior in realistic hardware reverse engineering scenarios.
To investigate if the twofold purpose of the projects was successful, a pilot course has been conducted \cite{tale:2018:wiesen}.
The results of the educational evaluation showed the successful implementation of the projects as educational and research materials.
Nevertheless the analysis revealed indications for revising aspects of the single projects.
In the following the revised projects which will be used for our upcoming larger studies are presented.

\subsubsection{Project 1 - Introduction to \HAL}
An introductory project helps students to understand the various features and the functionality of \HAL.
It imparts knowledge about different gate types and their functionalities, and how \HAL can be incorporated for the reverse engineering process.
Dedicated \HAL features like automated generation Boolean functions - or rather Binary Decision Diagrams (BDDs) - to analyze the logic of gates are introduced.
Moreover applying different algorithms from graph theory, e.g., identifying strongly connected components, are included in project 1, which will support students in solving hardware reverse engineering tasks.
For that matter a simple cipher has been designed consisting of a substitution-permutation-network.

\subsubsection{Project 2 - \ac{FSM} Reverse Engineering}
In this task an \ac{FSM} circuit has to be identified in \textit{the sea of gates} and the transition logic has to be reversed.
The challenge in this task is to find the \ac{FSM} in the netlist with several thousand gates.
The task makes heavy use of novel features of \HAL, like the submodules to group parts in the netlist that belong together, as described in \Section{co::sec::hal_improvements}.
The detection of the \ac{FSM} has become more realistic by finding false positive results when it comes to identifying possible \ac{FSM} circuits.
The challenge for the students is to analyze the candidates in more detail.
Algorithms from graph theory, supporting the search, can directly be executed via \HAL plugins.
Once the component has been identified the Boolean functions of the state transition logic has to be explored in order to retrieve the states and transitions.
Live exploration can be done easily with the integrated Python interface to control \HAL (cf. \Section{co::sec::hal_improvements}).

\paragraph{What psychological insights will we gain?}
\begin{itemize}
\item First reference point of how long students need to solve the \ac{FSM} reversing task in total.
\item Analysis and description of students' mistakes during the \ac{FSM} reversing task.
\item Analysis of how much extra time effort students spent on solving mistakes.
\end{itemize}

\subsubsection{Project 3 - Obfuscated \ac{FSM}}

\begin{figure}[htb]
	\centering
	\resizebox{0.7\linewidth}{!}{%
		\begin{tikzpicture}
		\tikzstyle{every state}=[fill=white,draw=black,text=black,inner sep=0pt,minimum size=17pt,circular drop shadow]

		\draw [help lines,red] (0,0) (8,5);

		\node[initial,state,fill=lightcoral] (sO0) at (2,2) {$s^\text{O}_0$};
		\node[state,fill=lightcoral] (sO1) at (3.5,3) {$s^\text{O}_1$};
		\node[state,fill=lightcoral] (sO2) at (5,2) {$s^\text{O}_2$};
		\node[state,fill=lightcoral] (sO3) at (3,1) {$s^\text{O}_3$};
		\node[state,fill=lightcoral] (sO4) at (4,1) {$s^\text{O}_4$};
		\node[state,fill=lightcoral] (sA0) at (1,4.5) {$s^\text{A}_0$};
		\node[state,fill=lightcoral] (sA1) at (3,4.5) {$s^\text{A}_1$};
		\node[state,fill=lightcoral] (sA2) at (5,4.5) {$s^\text{A}_2$};
		\node[state,fill=lightskyblue] (s0) at (7,1.25) {$s_0$};
		\node[state,fill=lightskyblue] (s1) at (7,2.25) {$s_1$};
		\node[state,fill=lightskyblue] (s2) at (7,3.25) {$s_2$};
		\node[state,fill=lightskyblue] (s3) at (7,4.25) {$s_3$};

		\path[->,>=stealth']
		(sO0) edge [above] node {$i_0$} (sO1)
		(sO0) edge (sA0)
		(sO1) edge [above] node {$i_1$} (sO2)
		(sO1) edge (sO3)
		(sO2) edge (sO4)
		(sO2) edge [anchor=center, above] node {$i_2$\ \ \ \ \ \ } (s0)
		(sO3) edge (sO0)
		(sO4) edge (sO3)
		(sO4) edge (sO0)
		(sO4) edge (sO1)
		(sO0) edge (sA0)
		(sA0) edge (sA1)
		(sA1) edge (sA2)
		(sA2) edge [bend right] (sO0)
		(s0) edge (s1)
		(s1) edge (s2)
		(s2) edge (s3)
		(s2) edge [loop left] (s2)
		(s3) edge [bend left] (s0);

		\draw [rounded corners, thick, dotted] (0.25,0.5) -- (5.5,0.5) -- (5.5,5) -- (0.25,5) -- cycle;
		\draw [rounded corners, thick, dashed] (6,0.5) -- (8,0.5) -- (8,5) -- (6,5) --cycle;

		\end{tikzpicture}
	}
	\caption{Obfuscated \ac{FSM} using HARPOON \cite{tcad:2009:chakraborty}}
	\label{co::fig::obfsucated_fsm}
\end{figure}
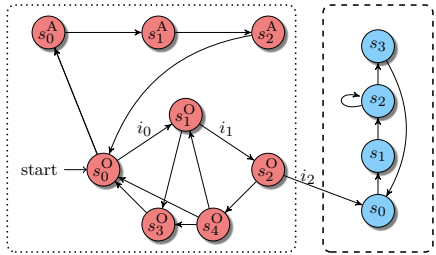

In \Section{asp-dac:chap:obfuscation} several obfuscation techniques to protect the control path - and thus the \ac{FSM} - have been described.
Therefore, Project 3 incorporates reversing of an \ac{FSM} based obfuscation technique named HARPOON \cite{tcad:2009:chakraborty}, as shown in \Figure{co::fig::obfsucated_fsm}.
HARPOON inserts a second \ac{FSM} (red) to protect the original \ac{FSM} (blue).
The inserted \ac{FSM} part has to be traversed in a certain way, using a specific input sequence, called the enabling key.
Every other input sequence than the enabling key will not lead to the original \ac{FSM} thus rendering the hardware design unusable for unauthorized parties.

In this project the students need to apply the methods they learned in Project 2 in order to detect the \ac{FSM} components in a gate-level netlist.
In addition, the challenge is to circumvent the implemented obfuscation HARPOON \cite{tcad:2009:chakraborty}.
In summary, two different attacks have to be performed:
First, the states and transitions of the \ac{FSM} have to be reversed to disclose the enabling key.
Second, students will learn how to reverse the obfuscation and manipulate gate-level netlists to patch the initial state of the \ac{FSM} \cite{ches:2018:fyrbiak}.

\paragraph{What psychological insights will we gain?}
\begin{itemize}
	\item Analysis of how much time students required to solve the second \ac{FSM} reversing task in total.
	\item Comparison with reference point of total required time in Project 2.
	\item Analysis if students have solved Project 3 more efficiently (e.g., shorter solution time, less mistakes, recovering from mistakes faster).
\end{itemize}

\subsubsection{Project 4 - AES}

Project 4 combines dynamic and static netlist reverse engineering by asking the students to extract a fixed key from an AES design.
Therefore the following steps have to be performed:
First the AES S-Box has to be identified in the netlist and manipulated using \HAL in order to weaken the AES \textit{(static part)}.
Secondly, the now manipulated netlist can be used for dynamic analysis.
Due to purposeful manipulation of the netlist, simulations of the design allow for extraction of the fixed key \cite{ivsw:2017:wallat}.

\paragraph{What psychological insights will we gain?}
\begin{itemize}
	\item Analysis and description of used hardware reverse engineering strategies in the most complex task of the study.
	\item Analysis and comparison of total solution time, made mistakes and time for solving mistakes.
	\item Descriptive analysis of how students recognized patterns and structures in finding the relevant candidates for extracting the AES key.
\end{itemize}


\section{Conclusion}
In this paper we highlighted serious security threats evolving from attacks based on hardware reverse engineering and identified the need for novel obfuscation methods impeding such attacks.
Based on the inseparable connection of technical aspects and human factors for conducting hardware reverse engineering we derived the necessity for obfuscation methods considering both technical aspects and the behavior of human analysts.
To adequately capture human reverse engineering strategies we significantly enhanced our hardware reverse engineering tool \HAL.
In the next step of our interdisciplinary research, we focused on the exploration of human factors in hardware reverse engineering by analyzing their learning processes and identifying common mistakes in four realistic scenarios.
Therefore, we methodically include both longitudinal and cross-sectional analyses to understand and characterize hardware reverse engineering processes of individuals at different levels of expertise.
Future studies will generate deeper insights about underlying human processes, which may allow for deduction of quantification metrics, eventually enabling better obfuscation methods.

    \section*{Acknowledgment}
	The research was supported in part by ERC Advanced Grant 695022 and NSF award NS-1563829.
	
	We also want to thank Sebastian Maassen and Adrian Drees for supporting the development of \HAL.

   \bibliographystyle{ACM-Reference-Format}
   \bibliography{bibliography}

\end{document}